\documentclass[aps,prl,twocolumn,superscriptaddress,floatfix]{revtex4-2}
\usepackage{amsfonts}
\usepackage{float}
\usepackage{booktabs} 
\usepackage{listings}
\usepackage{mathtools}
\usepackage{graphicx}
\usepackage{dcolumn}
\usepackage{bm}
\usepackage{amsmath}

\usepackage{ulem}
\usepackage{color}
\usepackage{amsthm}
\usepackage{breakcites}
\usepackage{textcomp}
\usepackage{multirow}
\usepackage{amssymb}
\usepackage[american]{babel}
\usepackage{braket}
\usepackage[breaklinks=true]{hyperref}
\usepackage{orcidlink}

\hypersetup{colorlinks=true,linkcolor=blue,citecolor=blue,filecolor=blue,urlcolor=blue,pdfstartview=FitH}

\begin{document}

\title{Rotation Breaks Kibble-Zurek Universality}
\author{Shiva Dwivedi\orcidlink{0000-0002-3803-0980}}
\email{shiva.dwivedi@christuniversity.in}
\affiliation{Centre for Quantum Technologies and Complex Systems (CQTCS), Christ University, Bengaluru, Karnataka-560074, India}
\author{Kenichi Kasamatsu\orcidlink{0000-0002-3825-9099}}
\email{kenichi@phys.kindai.ac.jp}
\affiliation{Department of Physics, Kindai University, Higashi-Osaka, Osaka 577-8502, Japan}
\author{Adolfo~del Campo\orcidlink{0000-0003-2219-2851}} 
\email{adolfo.delcampo@uni.lu}
\affiliation{Department of Physics and Materials Science, University of Luxembourg, L-1511 Luxembourg, Luxembourg}
\affiliation{Donostia International Physics Center, E-20018 San Sebasti\'an, Spain}

\author{Mithun Thudiyangal\orcidlink{0000-0003-4341-6439}}
\email{mithun.thudiyangal@christuniversity.in}
\affiliation{Centre for Quantum Technologies and Complex Systems (CQTCS), Christ University, Bengaluru, Karnataka-560074, India}

\begin{abstract}
We show that trap rotation overrides Kibble-Zurek (KZ) universality when a 2D Bose-Einstein condensate forms via a finite-time quench. In stochastic Gross-Pitaevskii simulations, the rotation-induced chiral bias drives the vortex number close to the Feynman-Onsager (FO) value, independent of the quench time. Ramping up rotation with the quench restores the KZ scaling at low frequencies, with vortex numbers below the FO value. The vortices form a disordered lattice that a pinning potential can order into an Abrikosov lattice, even at low rotation. Trap rotation thus offers a control knob for critical dynamics. 
\end{abstract}

\maketitle

When a continuous phase transition with $U(1)$ symmetry breaking is driven at a finite rate, its non-equilibrium dynamics is elegantly described by the Kibble-Zurek mechanism (KZM)~\cite{kibble1976topology, kibble1980some,Zurek85,Zurek96c,del2014universality}. The KZ framework relies on equilibrium properties near the critical point. Specifically, it invokes the divergence of the correlation length $\xi = \xi_0/|\epsilon|^\nu$ and the relaxation time $\tau = \tau_0/|\epsilon|^{z\nu}$, where $\nu$ and $z$ are the static and dynamic critical exponents, respectively, and $\epsilon$ is the dimensionless distance to the critical point of the parameter. The system's dynamics freezes out when the relaxation time exceeds the time remaining until the transition. As a result, the macroscopic response of the system is delayed by a characteristic equilibration time $\hat{t} = (\tau_0\tau_{\rm Q}^{z\nu})^{1/(1+z\nu)}$ that depends on the quench time $\tau_{\rm Q}$. This freeze-out establishes a characteristic correlation length  $\hat{\xi} = \xi_0 \left(\frac{\tau_{\rm Q}}{\tau_0}\right)^{\nu/(1+z\nu)}$. For point-like defects such as vortices, KZM predicts that defects form at a density $\rho_{\rm v}$ that scales universally with the quench rate as $\rho_{\rm v} \propto \frac{1}{\hat{\xi}^{D}} = \frac{1}{\xi_0^{D}}\left(\frac{\tau_0}{\tau_{\rm Q}}\right)^{D\nu/(1+z\nu)}$ where $D$ denotes the spatial dimension~\cite{Zurek85,Zurek96c}.

$U(1)$ symmetry breaking occurs naturally during the cooling of a Bose gas in finite time, resulting in the formation of a Bose-Einstein condensate (BEC)~\cite{PhysRev.54.947,anderson1995observation} that is spontaneously populated with topological defects~\cite{Weiler2008,delCampo2011}. The KZ universality has been confirmed in Bose gases~\cite{Navon2015,Goo21,Das12,Liu2020,zurek2009causality,thudiyangal2024universal} and has led to the phenomenon of spontaneous quantum turbulence~\cite{shinn2025spontaneous}, and the study of scaling laws for the circulation statistics ~\cite{Massaro2025}. Universality is also confirmed on a wide range of platforms, including superfluid $^3$He~\cite{ruutu1996vortex, bauerle1996laboratory} and atomic Fermi gases~\cite{Ko2019}, holographic superfluids and superconductors~\cite{Sonner15,Chesler15,ZengXia2023,Xia24,xia2026evolutionvortexstringsthermal}, ion chains~\cite{Nigmatullin16}, and multiferroics~\cite{Griffin12,Lin14}.
\begin{figure*}[t]
    \centering
    \includegraphics[width = 1.0\linewidth]{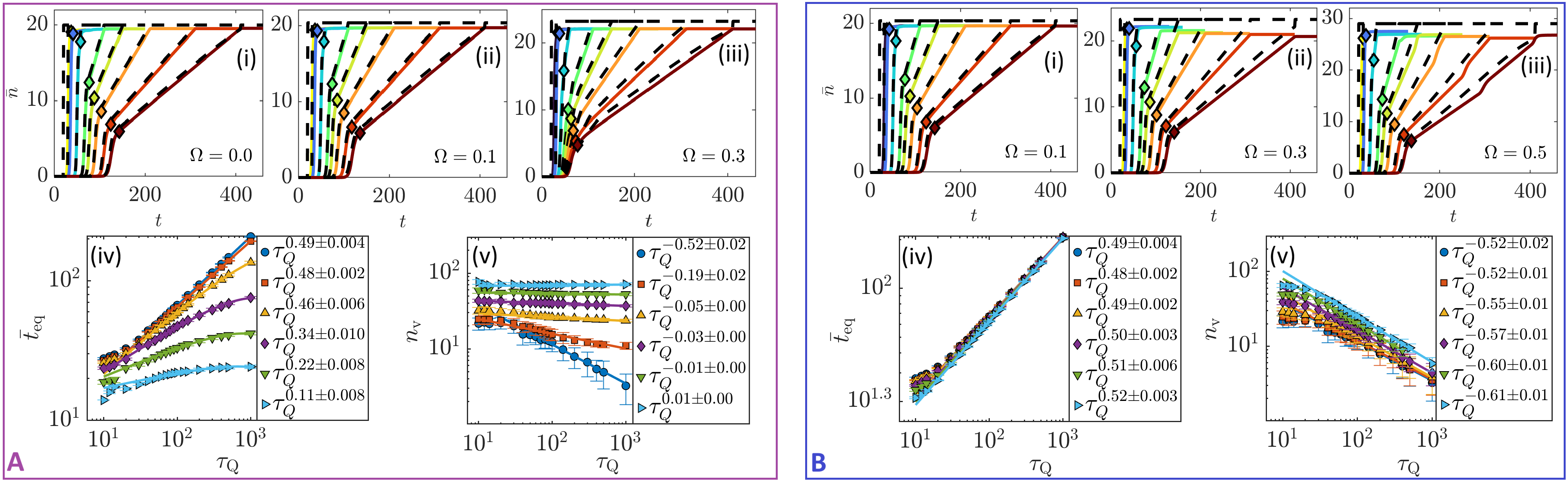}
    \caption{ \textbf{Response time of the order parameter in a BEC transition and KZ scaling} for a trapped condensate at [\textbf{A}] fixed rotation and for [\textbf{B}] a ramping rotational frequency. The norm of the condensate $\bar{n} = \frac{1}{A}\int_{A}|\phi (\textbf{r},t)|^2 d^2 r$ is plotted for $\tau_{\rm Q} \in \{10,20,50,100,200,300,400\}$ (left to right) at different rotational frequencies $\Omega = \text{0, 0.1, and 0.3}$, respectively in [\textbf{A}](i)-(iii) and ${\Omega_{\rm f}} = \text{0.1, 0.3, and 0.5}$, respectively in [\textbf{B}](i)-(iii). Dashed lines represent the analytically estimated norm of the condensate from Eq. \eqref{Eq: Analytical_Final density exression}. The diamonds on the growth curves correspond to the equilibration time $t_{\rm eq}$. [\textbf{A},\textbf{B}](iv) The crossover time $\bar{t}_{\rm eq} \mathrel{\mathop:}= (t_{\rm eq} - t_{\rm c})-t_{0}$, where $t_{\rm c}$ denotes the time at which the system crosses the critical point ($\mu_{\rm c}$), plotted as a function of the quench time $\tau_{\rm Q}$ on a log-log scale for $\Omega = \{ 0.1,0.2,\dots ,0.5 \}$ (top to bottom in the legends). [\textbf{A},\textbf{B}](v) Average vortex number calculated at $t_{\rm eq}$. The error bars in panels (iv) and (v) in both blocks correspond to the standard deviation from the mean across $\mathcal{R}=50$ different stochastic trajectories, while (i)-(iii) correspond to a single trajectory ($\mathcal{R}=1$).
    }
    \label{Fig:norm_evo_mu_and_scaling}
\end{figure*}

Experimenting with KZM in a rotating frame introduces an additional constraint from angular momentum that is absent from the conventional KZ picture. During the phase transition of Bose-Einstein condensation, a background rotational field can bias the spontaneous formation of defects and can lead to the survival of only vortices or antivortices depending on the sign of the angular momentum. As a result, deviations from the KZ universality are expected. On the other hand, a rotating condensate with enough angular momentum at equilibrium leads to the formation of an ordered array of vortices, generally known as the Abrikosov lattice~\cite{feynman1955application,fetter2009rotating, madison2000vortex,abo2001observation}. In this limit, the rotating condensate approaches rigid-body rotation with vortex number $n_{\rm v}$ dictated by the Feynman--Onsager (FO) relation $n_{\mathrm{FO}}\simeq\Omega R^{2}$~\cite{onsager1949statistical,feynman1955application,kasamatsu2003nonlinear}. This raises several intriguing questions: Do vortices that equilibrate following a finite-rate quench across a continuous phase transition obey the FO relation? How does the bias introduced by angular momentum affect the non-equilibrium generation of topological defects and the associated KZ universality? Furthermore, can the KZ mechanism facilitate the formation of an Abrikosov vortex lattice at low rotation frequencies, below the threshold for dynamical (surface-mode) vortex formation? 

In this Letter, we investigate the biases in the topological charges induced by angular momentum during Bose condensation in a rotating frame. We consider both condensation in a rotating trap at a fixed rotation frequency and condensation accompanied by a simultaneous ramp of the rotation frequency. By comparing the dynamics of the system during the two quenches, we establish that the trap rotation frequency at the crossover time affects the critical dynamics, breaking KZ universality. We further reveal that, after the quench, vortices biased by rotation settle into a lattice, which can then form an Abrikosov lattice in the presence of an external pinning potential.
\subparagraph*{Theoretical model:}
To describe the non-equilibrium condensate dynamics and the emergence of topological defects, we utilize the two-dimensional ($D=2$) stochastic Gross-Pitaevskii equation (SGPE) 
\begin{equation}
(i-\gamma)\frac{\partial \phi}{\partial t}= \left[\mathcal{H}_\Omega + g|\phi|^2 - \mu(t) \right]\phi + \sqrt{2\gamma T}\zeta(\mathbf{r},t),
\label{eq:gpe_compact}
\end{equation}
where $\mathcal{H}_\Omega = -\frac{1}{2}\nabla^2 - \Omega L_{\rm z} + V(\mathbf{r})$, $L_{\rm z} = -i({\rm x}\partial_{\rm y} - {\rm y}\partial_{\rm x})$ is the angular momentum operator along the ${\rm z}$-axis, and $g$ represents the coupling strength of the interaction between bosons. $V(\mathbf{r})$ is the external trapping potential, $\gamma$ denotes the phenomenological dissipation rate, and $T$ is the temperature of the thermal cloud. Thermal fluctuations are modeled by $\zeta(\mathbf{r},t)$, a complex Gaussian white-noise process with zero mean. Here, $\mu(t)$ is the chemical potential and is our control parameter. We consider a hard-wall potential given by $V(\mathbf{r}) = V_{\rm 0}\Theta( r^2 - R^2)$, where $\Theta(x)$ is the Heaviside step function. For illustration, we take $V_{\rm 0} = 120$ and set the radius of the trap $R = 12$. To quantify the macroscopic growth of the system, we monitor the norm of the condensate, $\bar{n}(t) = \frac{1}{A} \int_A |\phi(\mathbf{r},t)|^2 \, d^2r$, where $A$ is the total trap area, and $n(\mathbf{r},t)=|\phi(\mathbf{r},t)|^2$. We restrict our study to rotation frequencies up to $\Omega = 0.5$ to minimize vortex formation through other channels, including surface-mode instabilities~\cite{Dalfovo_Stringari_Shape_deformations, Sinha_Castin_Dynamical_Instability, Cai_yuan_critical_frequency, Anglin_surface_mode_spectrum,MT_critical_frequency_scatt_length}. Throughout, we use dimensionless units with $\hbar=k_{\rm B}=m=1$ (see~\cite{SM} for details and simulation parameters) \nocite{dennis2013xmds2, tkachenko1966vortex, abrikosov1957magnetic, kleiner1964bulk, abrikosov2004nobel, essmann1967direct, aragon2019unveiling, campbell1979vortex, stauffer1968distribution, K_Damle_1996, Giorgini_critical_temp_96, Theory_BEC_Dalfovo_Stefano_review_1999, Patra2026-ls}.

\subparagraph*{Case I: Condensation with fixed rotation frequency ($\mu$ quench):} 
We consider a linear quench of the chemical potential $\mu(t) = \mu_{\rm i} + (t-t_{\rm 0})v_\mu$, where $v_\mu = (\mu_{\rm f} - \mu_{\rm i}) /\tau_{\rm Q}$. Figure~\ref{Fig:norm_evo_mu_and_scaling}[A](i)-(iii) shows the temporal evolution of the norm for different quench times and rotation frequencies. The numerical growth curves match the analytical estimate of the norm 
well (see EM \ref{EM: Growth_Equation} for a detailed derivation),
\begin{equation}
n(\mathbf{r},t) = \frac{n_0 e^{A t^2 + B(\mathbf{r})t}} {1 + n_0 (2\gamma g) M(\mathbf{r},t)}.
 \label{Eq: Analytical_Final density exression}
\end{equation}
Here, $t$ is measured from the beginning of the quench, $n_0$ is the thermal seed density, and $A$, $B(\mathbf{r})$, and $M(\mathbf{r},t)$ are given in EM~\ref{EM: Growth_Equation}. During quench, the growth curve generally increases first exponentially and then linearly. This transition in the growth rate marks the equilibration time $t_{\rm eq}$ at which the number of defects are counted for probing the KZ scaling. The growth curve settles to the Thomas-Fermi limit once the quench ends.

\begin{figure*}[!t]
    \centering
    \includegraphics[width=1.0\linewidth]{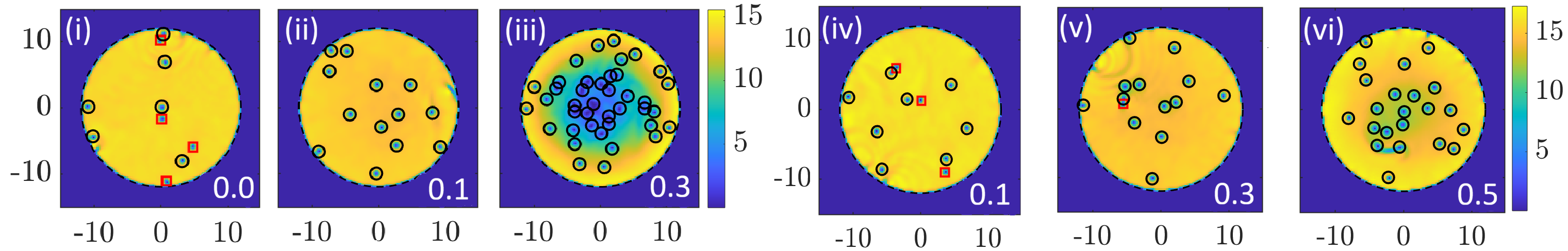}
    \caption{\textbf{Snapshots of the condensate at equilibration time $t_{\rm eq}$} (marked with diamonds in Fig. \ref{Fig:norm_evo_mu_and_scaling}) for a quench time $\tau_{\rm Q} = 100$ and different rotation frequencies, indicated on the lower right of each plot. Black circles identify co-rotating vortices, whereas the red squares mark counter-rotating vortices. Panels (i)-(iii) correspond to $\mu$-quenches and (iv)-(vi) to $(\mu,\Omega)$-quenches. The plots are shown for a single trajectory ($\mathcal{R} = 1$). The black dashed circle corresponds to $0.97R$, beyond which vortices are not counted.
    }
    \label{Fig:density_snapshots_mu_quench}
\end{figure*}
We calculate the crossover time, $\bar{t}_{\rm eq} \mathrel{\mathop:}= (t_{\rm eq} - t_{\rm c}) - t_0$, where $t_{\rm c}$ is the time at which the system crosses the critical point $\mu_{\rm c}$  (detailed in~\cite{SM}).  Thus, $\bar{t}_{\rm eq}$ is expected to be proportional to $\hat{t}$. The scaling exponent $\alpha$ can then be evaluated as $\bar{t}_{\rm eq}\propto \tau_{\rm Q}^\alpha$. In the mean-field regime ($z=2$, $\nu=1/2$), KZ theory predicts $\alpha \equiv z\nu/(1+z\nu)=1/2$. Averaging over $\mathcal{R}=50$ trajectories, we find $\alpha \approx 0.49\pm0.004$ in a non-rotating trap ($\Omega=0$) which is in good agreement with this prediction (Fig.~\ref{Fig:norm_evo_mu_and_scaling}[A](iv)). The deviation from the KZ prediction is observed at finite rotation, where $\bar{t}_{\rm eq}$ no longer follows the expected scaling with $\tau_{\rm Q}$, as seen in Fig.~\ref{Fig:norm_evo_mu_and_scaling}[A](iv).

As further detailed in EM \ref{EM_sec: Chirality}, one finds a finite chiral imbalance between co- and counter-rotating vortices at early times, $t\approx t_{\rm eq}$. Thus, by the time the condensate reaches the crossover from exponential to linear growth, its dynamics is already influenced by rotation. For slower quenches, the condensate has more time to acquire angular momentum and undergo rotationally induced radial density redistribution before the crossover. This enhances the density away from the trap center and causes the nonlinear interaction term $g|\phi|^2$ to become important earlier, deviating $\bar{t}_{\rm eq}$ from the conventional KZ scaling. At $\bar{t}_{\rm eq}$ the interaction energy, $g|\phi|^2$ attains a finite value, balancing the time-dependent chemical potential $\mu(t)$.

\subparagraph*{Case II: Condensation with ramping rotation frequency $((\mu,\Omega$) quench):}
We now start the quench with $\Omega_{\rm i}=0$ and ramp both $\mu$ and $\Omega$ together to $\mu_{\rm f}$ and $\Omega_{\rm f}$, respectively, over the same quench time $\tau_{\rm Q}$ (protocol details are given in~\cite{SM}). Figure~\ref{Fig:norm_evo_mu_and_scaling}[B](i)-(iii) shows the corresponding growth curves, and the fitted exponent $\alpha$ of the crossover time does not vary much with different $\Omega_{\rm f}$; see Fig.~\ref{Fig:norm_evo_mu_and_scaling}[B](iv). For larger $\tau_{\rm Q}$ (slow quench) the influence of rotation is marginal ($\Omega(t)\approx 0$) on the system near the critical point, and the quench effect is similar to the non-rotating case. On the other hand, for fast quenches $\Omega$ increases quickly and therefore rotation affects the crossover time $\bar{t}_{\rm eq}$. As a result, a small deviation in $\alpha$ is observed as $\Omega_{\rm f}$ increases, further suggesting that the relevant parameter that controls the modification of the KZ dynamics is not $\Omega_{\rm f}$, but rather $\Omega(t = \bar{t}_{\rm eq})$. The norm shows jump-like features at larger $\Omega_{\rm f}$ during the linear growth. This jump is due to the creation of additional vortices that enter the periphery by the transfer of angular momentum to the condensate (see EM~\ref{EMsec:jumps}).

\par

\subparagraph*{Vortex density scaling:}
KZ theory predicts the density of vortices formed spontaneously during the phase transition scales as $\rho_{\rm v} \propto \tau_{\rm Q}^{-\beta}$, where $\beta \equiv D\nu/(1+z\nu) \approx 0.5$ for $D=2$. For a 2D condensate, KZ scaling of the average vortex number $ \langle n_{\rm v} \rangle \propto \rho_{\rm v}$ at $t_{\rm eq}$ with the cooling rate has already been established in Ref.~\cite{thudiyangal2024universal}. In a non-rotating trap confined in a hard-wall potential, we further confirm this claim and find that $\beta \approx 0.52\pm 0.02$ for Case I and Case II as shown in Fig.~\ref{Fig:norm_evo_mu_and_scaling}[A-B](v). Our results reveal that the observed scaling for the average vortex number breaks down for $\Omega \neq 0$ in Case I, while for Case II KZ universality holds for smaller rotational frequencies ($\Omega_{\rm f} < 0.3$). In a non-rotating trap, both vortices and antivortices form in equal numbers to maintain a net zero angular momentum and annihilate by mutual collision or through losses at the trap boundary. By contrast, in the rotating frame, a fluctuation mode with angular-momentum quantum number $m$ has an excitation energy $\epsilon_m$ shifted as $\epsilon_m \to \epsilon_m - m\Omega$. Thus, the rotation can bias the growth of stochastic fluctuations with different angular momenta that already exist in the early stage of condensation. Indeed, by $t_{\rm eq}$,  co-rotating vortices outnumber counter-rotating ones, i.e., those rotating oppositely to the trap. Comparing Case I and Case II (Fig.~\ref{Fig:density_snapshots_mu_quench}(ii)--(iii) and (iv)--(v)) shows that the chiral bias is weaker in Case II, particularly at lower rotational frequencies.

\begin{figure*}[!t]
     \centering
     \includegraphics[width=1.0\linewidth]{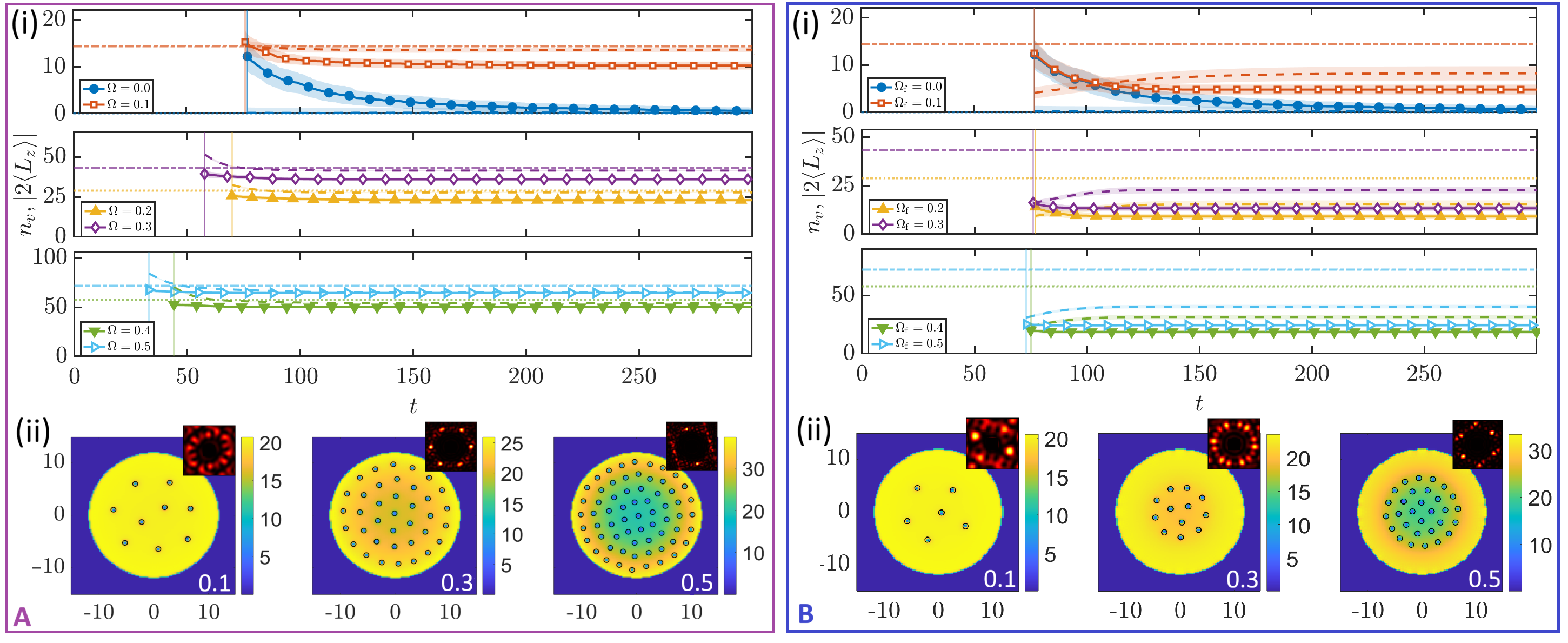}
     \caption{\textbf{(i) Emergence of rigid-body rotation and (ii) post-quench density profiles} for a trapped 2$D$ BEC at [\textbf{A}] fixed rotation and for [\textbf{B}] ramping rotational frequency. \textbf{(i)} Time evolution of the spontaneously formed vortex count $n_{\rm v}$ (solid lines) and scaled macroscopic angular momentum $2\langle L_{\rm z} \rangle$ (dashed lines) with a quench time $\tau_{\rm Q} = 100$, for \textbf{[A]} $\Omega \text{ and \textbf{[B]} } \Omega_{\rm f} \in \{0, 0.1, \dots, 0.5\}$. Vertical lines denote the equilibrium time $t_{\rm eq}$ (marked with solid diamonds in Fig. \ref{Fig:norm_evo_mu_and_scaling}[A-B](i)-(iii)). Horizontal dotted and dot-dashed lines mark the corresponding Feynman-Onsager (FO) limits, [A] $n_{\rm v} \approx \Omega R^2$ and [B] $n_{\rm v} \approx \Omega_{\rm f} R^2$. The plots are averaged over $\mathcal{R} = 50$ trajectories. The shaded regions show the standard deviation from the mean across $50$ trajectories. \textbf{(ii)} Post quench $2D$ density profiles for $\tau_{\rm Q} = 100$, [A]  $\Omega~\text{and}~ [B]~\Omega_{\rm f} \in \{0.1, 0.3, 0.5\}$ at $t = 300$. The black circles mark co-rotating vortices. The corresponding vortex-position structure factor $S(\mathbf{k}) = \frac{1}{n_v} \left| \sum_{j=1}^{n_v} e^{-i\mathbf{k}\cdot\mathbf{r}_j} \right|^2$ is displayed in the inset.
     }
     \label{Fig:angular momentum}
 \end{figure*}

\subparagraph*{FO limit and post-equilibration ($t>t_{\rm eq}$) dynamics:}
For a rotating superfluid, the macroscopic rotational field governs the equilibrium state of the system and requires a minimum number of vortices, which carry quantized angular momentum, to match the global angular momentum of the trap, $n_{\rm v} \approx 2\langle L_{\rm z}\rangle$~\cite{feynman1955application,onsager1949statistical,fetter2009rotating}. This requirement for a circular trap reduces to $n_{\rm FO} = \Omega R^2$~\cite{kasamatsu2003nonlinear, Adhikari_2019}. Therefore, during the condensation in a trap with finite angular momentum, the number of spontaneously formed vortices can be constrained by the FO limit.
To probe this scenario, we track the temporal evolution of the vortex count $n_{\rm v}(t)$ and the macroscopic angular momentum per particle $\langle L_{\rm z}(t) \rangle=\int \phi^*L_{\rm z}\phi\, d^2r/\int|\phi|^2 d^2r$ from the equilibration time. This gives us insights into the following questions: Which dynamics is governing the vortex formation? When and where exactly is the KZ defect formation violated? For $\Omega = 0$ it is expected that $\langle L_{\rm z} \rangle \approx 0$ and is confirmed in Fig.~\ref{Fig:angular momentum}[A-B](i). 
 
For $\Omega = 0.1$ Fig.~\ref{Fig:angular momentum}[A](i) indicates that $n_{\rm v}<|2\langle L_{\rm z}\rangle|<\Omega R^2$ for all $t>t_{\rm eq}$. This difference results from the interplay between the KZM and the macroscopic requirement of a rotating condensate. For $\Omega > 0.1$, we see that the system approaches the FO limit at an early time, $t \sim t_{\rm eq}$. The saturated vortex number remains slightly below $\Omega R^2$, as a result of the annihilation of co- and counter-rotating vortices before the equilibration time $t< t_{\rm eq}$. This crossover to a dynamics constrained by rotation overrides the universal KZ scaling and is the reason for the flattened power-law exponents observed in Fig.~\ref{Fig:norm_evo_mu_and_scaling}[A](v).

In contrast to Case I, Fig.~\ref{Fig:angular momentum}[B](i) shows that, for Case II, the system does not reach the FO limit even at late times ($t\gg\tau_{\rm Q}$). Here, as confirmed in EM~\ref{EM_sec: Chirality}, the rotational bias remains weak near the critical point, leading to the formation of comparable numbers of co- and counter-rotating vortices as in a non-rotating trap. Their subsequent annihilation via mutual collisions significantly reduces their density in the bulk. This is the dominant decay mechanism in this case and explains the observed steepening of exponents in Fig.~\ref{Fig:norm_evo_mu_and_scaling}[B](v). These losses due to mutual collision further restrict the system from attaining the macroscopic requirement of a rotating condensate.

It is natural to address what equilibrium configuration emerges for $t\gg \tau_{\rm Q}$. Figure~\ref{Fig:angular momentum}[A-B](ii) shows that after the quench, vortices settle in a lattice configuration. However, the sensitivity of the vortex number to the chiral bias and constraints from the trap geometry prevent an Abrikosov vortex lattice arrangement in the system. We identify the emergent lattice geometry as that of a vortex-glass phase, detailed in~\cite{SM}. In EM~\ref{EMsec: Emergence of Order via Pinning}, we further show that an Abrikosov lattice can be obtained at low rotation frequencies under a continuous external pinning potential. Nevertheless, removal of the pinning potential brings the lattice back to its original disordered arrangement.

\par

KZ scaling has been observed directly in experiments with Bose gases, in both non-uniform traps~\cite{Lamporesi2013,Goo21} and uniform box traps~\cite{Navon2015}. Further, similar uniform (hard-wall) trap geometries have been realized experimentally, from the earliest optical box traps~\cite{meyrath2005box} to their modern refinements~\cite{guant2013, Navon2021QuantumGases} including planar, disc-shaped box potentials realized in Ref.~\cite{johnstone2019flow}. On the other hand, rotating harmonic traps, realized by stirring the condensate with a focused laser beam or a rotating anisotropic potential, have long been used to form vortices and vortex lattices in BECs~\cite{madison2000vortex, abo2001observation}, and condensation directly inside an already-rotating trap, as required for our Case I, has also been demonstrated by evaporatively spinning up a thermal cloud before it crosses the condensation threshold~\cite{haljan2001driving}. Following these established techniques, combining a rotating box trap with a controlled ramp of the rotation frequency during the quench would allow a direct experimental test of both protocols studied in this work.

\paragraph{Discussions and Conclusion:} In summary, we have shown that during Bose-Einstein condensation, trap rotation plays an important role in the symmetry breaking process. In a rotating trap, the centrifugal force affects the freeze-out time scaling predicted with KZM. Trap rotation biases the formation of vortices towards the co-rotating vortices bringing the number close to the FO limit, and the universal KZ scaling predictions for the defect density are lost. By comparing two different quench protocols, we reveal that this modification of KZ dynamics is governed by the angular momentum experienced by the system during the early stage of dynamics. Finally, we show that the co-rotating vortices settle into a disordered lattice, which a continuous external pinning potential can order into an Abrikosov lattice, even at low rotation frequencies. Our findings reveal an interesting interplay between the stochastic process that leads to the expected universal KZ scaling behavior, and the macroscopic requirement from the global angular momentum that suppresses the expected scaling in the system. More generally, rotation is one instance of a background field that favors defects of one charge over their antidefects, and we anticipate that such fields can serve as control knobs for critical dynamics across the homotopy classification of phase transitions.

\begin{acknowledgments}
\paragraph{Acknowledgments:}
 MT thanks the Anusandhan National Research Foundation (ANRF), Government of India, for financial support through the Prime Minister Early Career Research Grant with Grant No. ANRF/ECRG/2024/003150/PMS, and Christ University for funding the research through the
seed grant, Sanction No.CU-ORS-SM-24/94. AdC acknowledges financial support from the Luxembourg National Research Fund under Grants No. C22/MS/17132060/BeyondKZM. 
\end{acknowledgments}

\bibliographystyle{apsrev4}
\bibliography{references}

\clearpage
\newpage

\onecolumngrid
\section{End Matter}

\twocolumngrid

\makeatletter
\renewcommand{\thesubsection}{\Roman{subsection}}
\def\@seccntformat#1{EM \csname the#1\endcsname:\ }
\makeatother
\setcounter{secnumdepth}{3}

\setcounter{equation}{0}
\renewcommand{\theequation}{A\arabic{equation}}

\subsection{Growth equation \label{EM: Growth_Equation}}

Recently, Suzuki and Zurek~\cite{suzuki2025deconstructing} presented analytical solutions of the Langevin equation and the GPE in a time-dependent Ginzburg-Landau potential for the non-equilibrium critical dynamics. The solution identifies the presence of $\pm \hat{t}$, the effective nonequilibrium response time. The authors focus on the universal early growth of the order parameter as a complementary probe to observing the scaling of the emerging defects. Our results show a good match with the analytical solution ( Eq.~\eqref{AppEq: Final density exression}) even at finite rotation, where no KZM scaling is found. Therefore, we conclude that mere monitoring and matching of the growth curve of the order parameter with the analytical predictions does not validate KZ scaling. 

The rotation effect can be applied to the Hamiltonian as $\mathcal{H}_\Omega = \frac{1}{2}(\mathbf{p} - \mathbf{A})^2 + V_{\rm eff}(\mathbf{r})$, where $\mathbf{A} = \boldsymbol{\Omega} \times \mathbf{r}$ acts as a synthetic vector potential analogous to that of a magnetic field acting on charged particles. The remaining spatial terms form the effective potential in the rotating frame: $V_{\rm eff}(\mathbf{r}) = V(\mathbf{r}) - \frac{1}{2}(\boldsymbol{\Omega} \times \mathbf{r})^2.$ For the circular $2D$ hard-wall trap,
\begin{equation} 
V_{\rm eff}(\mathbf{r}) = V(\mathbf{r}) - \frac{1}{2}\Omega^2 r^2. 
\end{equation} 
Inside the hard-wall trap, $V(\mathbf{r}) = 0$. Therefore, the effective potential is purely negative: $V_{\rm eff} = -\frac{1}{2}\Omega^2 r^2$, $r^2 = {\rm x}^2+{\rm y}^2$. This has an inverse parabola-like feature because of the centrifugal force. The Thomas-Fermi density is then given by $n = (\mu(t) - V_{\rm eff}) / g$:
\begin{equation} 
n(\textbf{r}) = \frac{\mu(t) - \left(-\frac{1}{2}\Omega^2 r^2\right)}{g} = \frac{\mu(t) + \frac{1}{2}\Omega^2 r^2}{g}.
\label{Appeq: effective chemical potential}
\end{equation}
Following the approach in~\cite{suzuki2025deconstructing}, we derive an analytical expression in the presence of rotation. From the dissipative Gross-Pitaevskii equation (SGPE without noise), the density evolution can be given as

\begin{equation}
\frac{\partial n}{\partial t}
= 2\gamma \left[ \left(\mu(t) + \frac{1}{2}\Omega^2 r^2\right)n - g n^2 \right].
\end{equation}

Solving this equation with $n(\mathbf{r},0)=n_0$, where $t$ is measured from the start of the quench, provides the density evolution in the presence of rotation, as

\begin{equation}
n(\textbf{r},t) = \frac{n_0 e^{A t^2 + B(\textbf{r})t}} {1 + n_0 (2\gamma g) M(\textbf{r},t)},
\label{AppEq: Final density exression}
\end{equation}

here $A = \frac{\gamma (\mu_{\rm f} - \mu_{\rm i})}{\tau_{\rm Q}}$, and $B(\textbf{r}) = 2\gamma \mu_{\rm i} + \gamma \Omega^2 r^2$, and

\begin{equation}
M(\textbf{r},t) = \frac{\sqrt{\pi} e^{-\frac{B(\textbf{r})^2}{4A}}}{2\sqrt{A}} \left[ \operatorname{erfi}\!\left(\sqrt{A}t + \frac{B(\textbf{r})}{2\sqrt{A}}\right)- \operatorname{erfi}\!\left(\frac{B(\textbf{r})}{2\sqrt{A}}\right)\right].
\end{equation}
For the $(\mu,\Omega)$ quench, the term $\gamma\Omega^2r^2t$ in the exponent in Eq.~\eqref{AppEq: Final density exression} is replaced by $\gamma r^2\int_0^{t}\Omega^2(s)\,ds$, and the solution is obtained numerically~\cite{SM}.

\subsection{Chirality measurement during condensation \label{EM_sec: Chirality}}

We now quantify the difference in vortices $|n_+ - n_-|$ with respect to the total vortex number $n_+ + n_-$ where $n_+$ and $n_-$ denote the number of co-rotating and counter-rotating vortices, respectively. We plot the quantity $\frac{|n_+ - n_-|}{n_+ + n_-}$ from the beginning of condensation ($t =t-t_0$) to the equilibration time, $t_{\rm eq}$. Here, $|n_+ - n_-|$ represents the net vorticity and measures the chiral bias induced by the angular momentum. 

\begin{figure}[htbp]
    \centering
    \includegraphics[width=1.0\columnwidth]{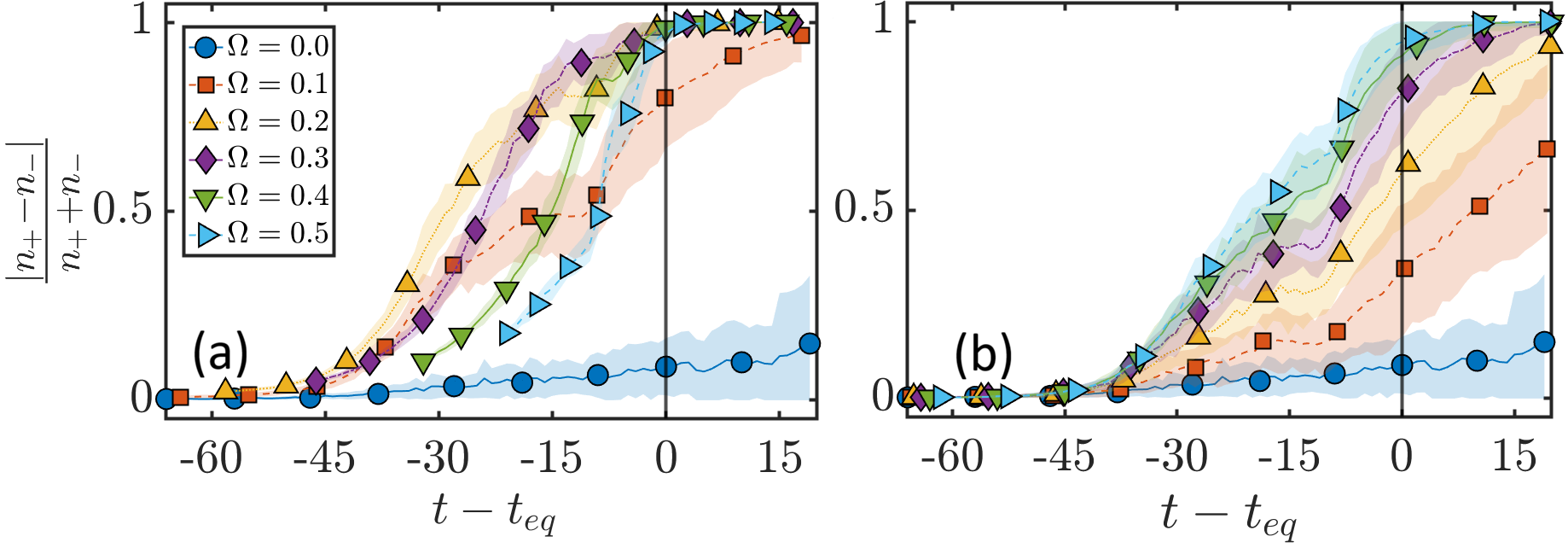}
    
    \caption{\textbf{Chiral bias in a rotating trap.} $\frac{|n_+ -n_-|}{n_+ + n_-}$ is plotted for $t<t_{\rm eq}$ for $\Omega \in \{0,0.1,\dots , 0.5\}$ ($\Omega_{\rm f}$ in (b)) for quench time $\tau_{\rm Q} = 100$ from the average of $\mathcal{R} = 50$ trajectories. The shaded regions show the standard deviation from the mean across $50$ trajectories. \textbf{(a)} Corresponds to $\mu$-quenches whereas \textbf{(b)} corresponds to ($\mu,\Omega$)-quenches.}
    \label{fig:polarization}
\end{figure}

\begin{figure*}[t]
    \centering
    \includegraphics[width=\linewidth]{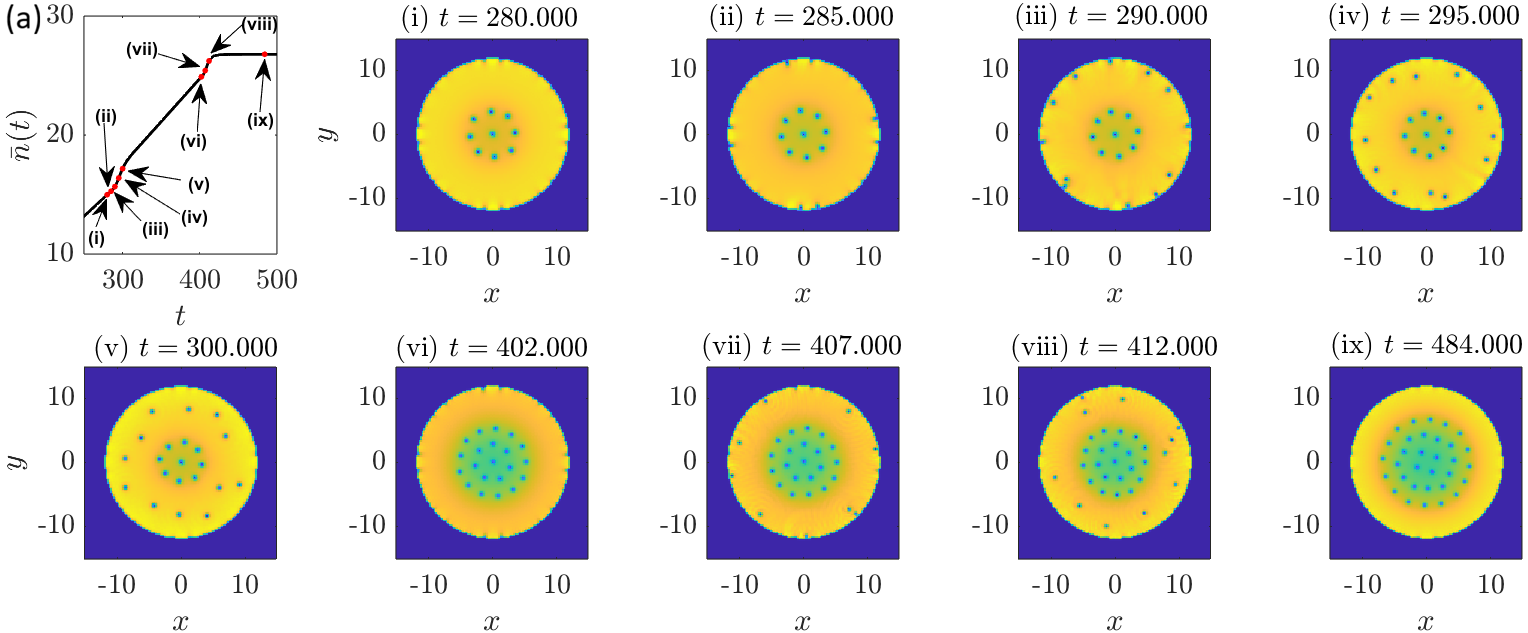}
    \caption{\textbf{Jumps in the norm of ($\mu,\Omega$)-quenches at larger $\tau_{\rm Q}$.} \textbf{(a)} Jumps in norm of the condensate for $\Omega_{\rm f} = 0.5$, $\tau_{\rm Q} = 400$. \textbf{(i)-(ix)}  density profile of the $2D$ BEC at the times of evolution highlighted in (a).}
    \label{fig:jump}
\end{figure*}

In Case I, for a strong angular momentum, Figure \ref{fig:polarization}(a) shows that $\frac{|n_+ - n_-|}{n_+ + n_-}(t \ll t_{\rm eq}) \to 1$ and implies that co-rotating vortices ($n_+$) lead the counter-rotating vortices ($n_+ \gg n_-$) from the onset of the quench ($t \ll t_{\rm eq}$). This suggests that the formation mechanism is highly affected by the angular momentum in the trap and plays a significant role during condensation, thus driving the vortex count toward the FO limit, $n_{\rm v} \to n_{\rm FO}$. In contrast, $\frac{|n_+ - n_-|}{n_+ + n_-} (t\ll t_{\rm eq}) \to 0$ for Case II in Fig.~\ref{fig:polarization}(b) implies that a comparable number of co- and counter-rotating vortices form ($n_+ \approx n_-$) and the effect of angular momentum is minimal on the vortex formation mechanism in this case.

\subsection{The jump behavior in the norm of $(\mu,\Omega)$-quenches for larger $\tau_{\rm Q}$}
\label{EMsec:jumps}
The norm evolution for $\Omega_{\rm f} = 0.5$, $\tau_{\rm Q}=400$ shows sudden jumps during linear growth, displayed in Fig.~\ref{fig:jump}(a). We notice that this jump is due to the creation of additional vortices entering through the periphery from the transfer of angular momentum to the condensate. This can be seen in Fig.~\ref{fig:jump}(i)--(ix). These vortices are not spontaneously formed  and demonstrate the ongoing interplay between stochastic defect formation and the macroscopic angular momentum.

\subsection{Emergence of order via pinning} \label{EMsec: Emergence of Order via Pinning}

In Fig. \ref{Fig:angular momentum}, the structure factor of the obtained vortex lattice exhibits diffuse ring-like features that demonstrate the lack of discrete rotational or translational symmetry. We next show that an optical pinning potential can be utilized to drive this vortex lattice~\cite{Pinning_potential_Phase_diagram_2004_Rejinder, Structural_PT_in_optical_lattices_2005_Bigelow,Pinning_observation_Cornell_2006} to a well-ordered Abrikosov lattice arrangement. We use a pinning potential of the form used in Ref.~\cite{sato2007vortex}: 
\begin{equation}
    V_{\text{pin}}(\mathbf{r}) = \sum_{n_1,n_2\in\mathbb{Z}} V_{\rm 0L} \exp\left[ - \frac{|\mathbf{r}-\mathbf{r}_{n_1,n_2}|^2}{(\sigma /2)^2} \right],
\end{equation}
where $\mathbf{r}_{n_1,n_2}= n_1 \mathbf{a}_1 + n_2 \mathbf{a}_2$. $V_{\rm 0L}$ and $\sigma$ denote the pinning amplitude and beam width, respectively. We employ a triangular geometry characterized by lattice vectors $\mathbf{a}_1 = a(1,0)$ and $\mathbf{a}_2 = a(-1/2, \sqrt{3}/2)$. The lattice constant $a$ is dictated by the rotational frequency $ a(\Omega) = \sqrt{\frac{2\pi}{\sqrt{3}\Omega}} $~\cite{sato2007vortex}. The pinning potential is applied instantaneously at the end of the quench and is maintained thereafter.

\begin{figure}[H]
    \includegraphics[width =1.0\linewidth]{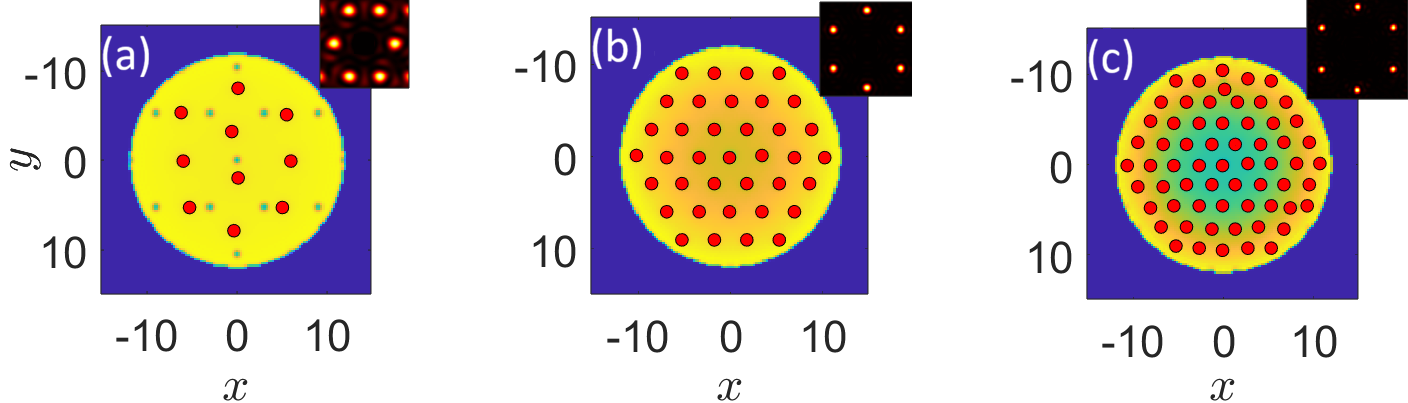}
    
    \caption{\textbf{Effect of pinning potential.} Density profiles of the $2D$ BEC after the quench in presence of a pinning potential for (a) $\Omega = 0.1$, (b) $\Omega = 0.3$ and (c) $\Omega = 0.5$ presented in Fig. \ref{Fig:angular momentum}[A](ii). For $\Omega = 0.1$, we let the system relax till $t = 1500$ and present the snapshot whereas for $\Omega = 0.3, ~0.5$ snapshot is taken at $t = 210$. The insets show their corresponding vortex-position structure factor $S(\textbf{k})$. Pinning parameters: beam width $\sigma = 0.7$, (a) $V_{\rm 0L} =15, ~ a = 6.022$, (b) $V_{\rm 0L} = 5, ~ a = 3.47$, and (c) $V_{\rm 0L} = 5, ~ a = 2.69$. }
    \label{fig: Effect of pinning potential}
\end{figure}

Under the pinning potential $V_{\text{pin}}$, sharply defined Bragg peaks emerge in $S(\mathbf{k})$, confirming the establishment of a long-range hexagonal order. It is seen that upon removal of the pinning potential, the vortex lattice returns to its original arrangement. Figure \ref{fig: Effect of pinning potential} presents the lattice arrangement of Case I, and we have verified a similar behavior for the lattices resulting from Case II. 

\end{document}


\title{---Supplementary Material---\\ 
Rotation Breaks Kibble-Zurek Universality}
\author{Shiva Dwivedi\orcidlink{0000-0002-3803-0980}}
\email{shiva.dwivedi@christuniversity.in}
\affiliation{Centre for Quantum Technologies and Complex Systems (CQTCS), Christ University, Bengaluru, Karnataka-560074, India}
\author{Kenichi Kasamatsu}
\email{kenichi@phys.kindai.ac.jp}
\affiliation{Department of Physics, Kindai University, Higashi-Osaka, Osaka 577-8502, Japan}
\author{Adolfo~del Campo\orcidlink{0000-0003-2219-2851}} 
\email{adolfo.delcampo@uni.lu}
\affiliation{Department  of  Physics  and  Materials  Science,  University  of  Luxembourg,  L-1511  Luxembourg, Luxembourg}
\affiliation{Donostia International Physics Center,  E-20018 San Sebasti\'an, Spain}

\author{Mithun Thudiyangal\orcidlink{0000-0003-4341-6439}}
\email{mithun.thudiyangal@christuniversity.in}
\affiliation{Centre for Quantum Technologies and Complex Systems (CQTCS), Christ University, Bengaluru, Karnataka-560074, India}

\maketitle

\makeatletter
\renewcommand{\thesection}{\Roman{section}}
\def\@seccntformat#1{SM \csname the#1\endcsname.\ }
\makeatother
\setcounter{secnumdepth}{3}
\renewcommand{\thefigure}{S\arabic{figure}}
\renewcommand{\theequation}{S\arabic{equation}}

\section{Simulation parameters and details}\label{Appsec:Simulation_parameters_and_details}
We consider the 2D stochastic Gross-Pitaevskii equation (SGPE),
\begin{equation}
(i-\gamma)\frac{\partial \phi}{\partial t}= \left[\mathcal{H}_\Omega + g|\phi|^2 - \mu(t) \right]\phi + \sqrt{2\gamma T}\zeta(\mathbf{r},t),
\label{eq:gpe_compact}
\end{equation}
where $\mathcal{H}_\Omega = -\frac{1}{2}\nabla^2 - \Omega L_{\rm z} + V(\mathbf{r})$. $L_{\rm z} = -i({\rm x}\partial_{\rm y} - {\rm y}\partial_{\rm x})$ is the angular momentum operator along the ${\rm z}$-axis, $g$ is the coupling strength of the delta-function interaction between bosons, $V(\mathbf{r})$ is the external trapping potential, $\gamma$ is the phenomenological dissipation rate, and $T$ is the temperature of the thermal cloud. We set $\hbar = k_{\rm B} = m = 1$. Thermal fluctuations are modeled by $\zeta(\textbf{r},t),$ a complex-valued Gaussian white-noise field with zero mean, satisfying $\langle \zeta(\textbf{r},t) \zeta^*(\textbf{r}',t')\rangle= \delta_{\textbf{r},\textbf{r}'}\delta_{t,t'}$. Equation~\eqref{eq:gpe_compact} is rendered dimensionless in length, time and energy by normalizing with the healing length $\xi_{\rm h} = \sqrt{\hbar^2/(m g n_0)}$, $m\xi_{\rm h}^2/\hbar$, and $\hbar^2/(m\xi_{\rm h}^2)$, respectively, where $m$ is the atomic mass and $n_0$ is the initial density in the trap $(V = 0)$. Temperature is measured in units of $\hbar^2/(k_B m \xi_{\rm h}^2)$.  

\par

To drive the phase transition, we initialize the system with a vanishing order parameter ($\phi = 0$) and allow it to relax in the presence of noise for a period $t_{\rm 0} = 10$. During the period $t_{\rm 0}$, $\mu(t)$ is held constant at $\mu_{\rm i}$, this ensures a vanishing order parameter. After $t_{\rm 0}$, we begin the quench as $\mu(t) = \mu_{\rm i} + (t-t_{\rm 0})v_\mu$, where $v_\mu = (\mu_{\rm f} - \mu_{\rm i}) /\tau_{\rm Q}$ and let it relax after the quench is over. We consider the temperature of the system $T=10^{-6}$, $\mu_{\rm i} = -0.1$, $\mu_{\rm f} = 20.0$, $\gamma = 0.03$, and $g = 1$. Simulations are performed with the XMDS2 software package \cite{dennis2013xmds2} using a semi-implicit integration algorithm. We simulate the system on an $N \times N$ spatial grid with $N=128$, where both the $x$ and $y$ dimensions span from $-15$ to $15$.
\par
In the $(\mu,\Omega)$-quench protocol, we quench $\mu$ and $\Omega$ from an initial value of $\mu_{\rm i}$ and $\Omega_{\rm i}$, respectively. Similarly to the previous protocol, we let the system relax for $t_{\rm 0} = 10$ at the initial values of the variables. Following $t_{\rm 0}$, we rotate the trap from $\Omega_{\rm i} = 0.0$ to $\Omega_{\rm f}$ linearly during a given quench time $\tau_{\rm Q}$ as $\Omega(t) = \Omega_{\rm i} + (t-t_{\rm 0})v_\Omega$, where $v_\Omega = (\Omega_{\rm f} -\Omega_{\rm i})/\tau_{\rm Q}$. After the quench, we set $\Omega(t) = \Omega_{\rm f}$ similar to $\mu(t) = \mu_{\rm f}$.

\section{Numerical calculation of $n(\mathbf{r},t)$}
\label{Appsec:Numerical_Calcualtion_of_nr,t}
The entire physical process is driven by the time-dependent chemical potential $\mu(t)$, which brings the system through a phase transition. This driving schedule involves three steps: (i) Pre-quench ($t \le t_{\rm 0}$): $\mu(t) = \mu_{\rm i}$, (ii) Actual quench ($t_{\rm 0} < t \le t_{\rm 0} + \tau_{\rm Q}$): $\mu(t) = \mu_{\rm i} + (t-t_{\rm 0})v_\mu$, and (iii)  Post-quench ($t > t_{\rm 0} + \tau_{\rm Q}$): $\mu(t) = \mu_{\rm f}$. The corresponding numerics are implemented in three different phases. \par
\textit{Phase A: The Thermal Seed ($t \le t_{\rm 0}$).} Before the quench, the system is initialized in the thermal phase with $|\phi| = 0$. In this phase, we model the thermal fluctuations via a complex Gaussian white noise $\zeta (\mathbf{r},t)$ with zero mean. The initial density at this stage is modeled based on the temperature $T$ and dissipation $\gamma$:~$n_{\rm ana}(\mathbf{r},t \leq t_0) = n_0 = 2\gamma T$. \par

\textit{Phase B: The Quench Dynamics ($t_{\rm 0} < t \le t_{\rm 0} + \tau_{\rm Q}$).} The exact analytical solution to the logistic growth equation during the quench is: $$n_{\rm ana}(\textbf{r},t) = \frac{n_0 e^{A t^2 + B(\textbf{r})t}} {1 + n_0 (2\gamma g) M(\textbf{r},t)},$$ with $A$, $B(\mathbf{r})$, and $M(\mathbf{r},t)$ given in EM I of the main text. For the $(\mu,\Omega)$ quench, the underlying logistic equation [Eq.~(A3) of the main text] remains valid with $\Omega\to\Omega(t)$, the term $\gamma\Omega^2r^2t$ in the exponent in [Eq.~(A4) of the main text] is replaced by $\gamma r^2\int_{t_0}^{t}\Omega^2(s)\,ds$, and the closed form above no longer applies. In both the cases, the logistic equation is integrated numerically with a fourth-order Runge-Kutta (RK4) scheme.

\textit{Phase C: The Thomas-Fermi Plateau ($t > t_{\rm 0} + \tau_{\rm Q}$).}  Once the quench is completed and the system relaxes, the derivative $\frac{\partial n}{\partial t} \rightarrow 0$. The logistic equation naturally reduces to the standard inverted-parabola Thomas-Fermi profile with rotation: $n_{\rm ana}(r) = \frac{\mu_{\rm f} + \frac{1}{2}\Omega^2 r^2}{g}.$
\par
To collapse the $2D$ grid into a $1D$ curve $n(r)$, we average the density inside thin concentric rings. For a ring at radius $r_{\rm i}$ with thickness $\Delta r$: $n(r_{\rm i}) = \frac{1}{2\pi r_{\rm i} \Delta r} \int_{\text{shell}} n({\rm x},{\rm y}) d{\rm x} d{\rm y}.$ To this end, we track the condensation over time by integrating the density over the entire $2D$ trap area $A$ and dividing by $A$, which gives the norm: $\bar{n}(t) = \frac{1}{A} \int_{{\rm x}^2 + {\rm y}^2 \le R^2} n({\rm x},{\rm y},t) d{\rm x} d{\rm y}.$

\begin{figure}[!h]
    \centering
    \includegraphics[width=1.0\linewidth]{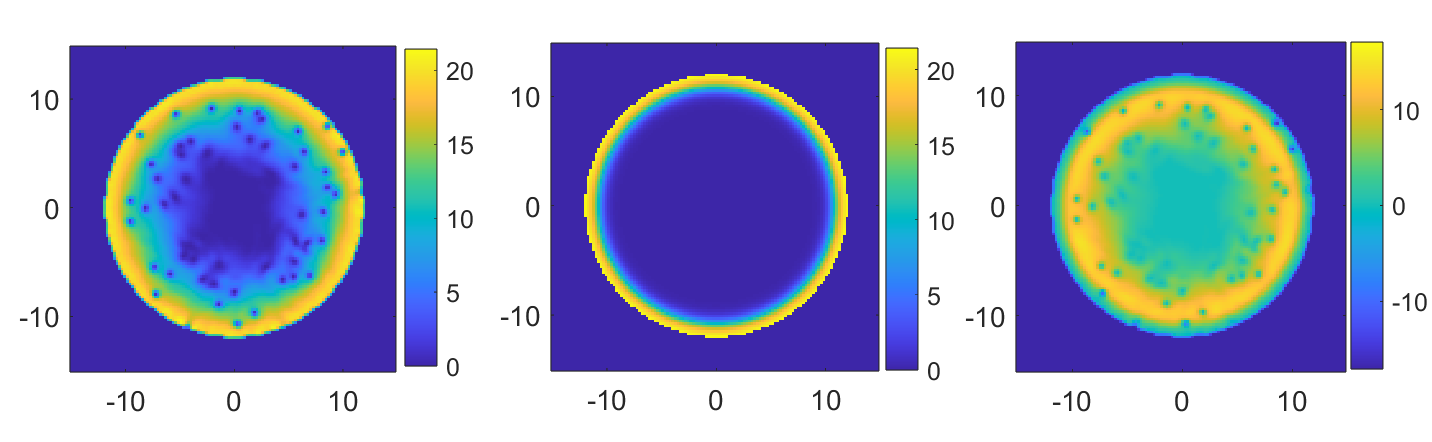}
    \includegraphics[width=1.0\linewidth]{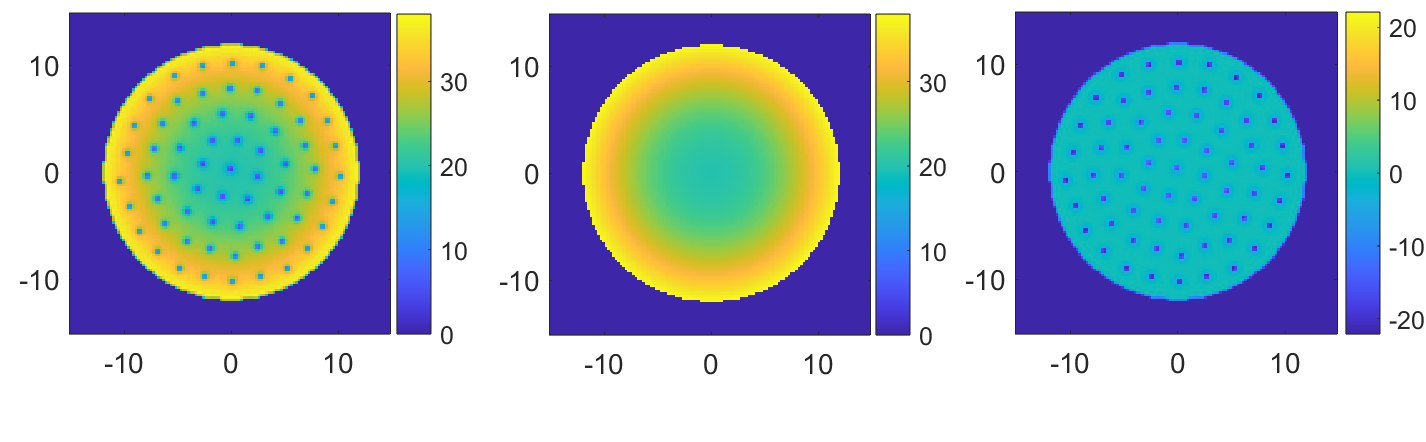}
    \caption{\textbf{Numerical data and analytical solution for order parameter growth during condensation.} The comparison of $n(\textbf{r},t)$ between the numerical data for $\mathcal{R} =1$ and the analytical solution is presented for $\Omega = 0.5$, $\tau_{\rm Q} = 100$ for (top row) $t=30$ and (bottom row) $t = 200$. The left and middle columns present the numerical and analytical data, respectively. The right column plots the difference between the corresponding values.}
    \label{Appfig:numerical_vs_analytical_comparison}
\end{figure}

\section{Vortex Lattice}
\label{Appsec:Vortex_lattice}

\begin{figure}[t]

    \includegraphics[width = 1.0\linewidth]{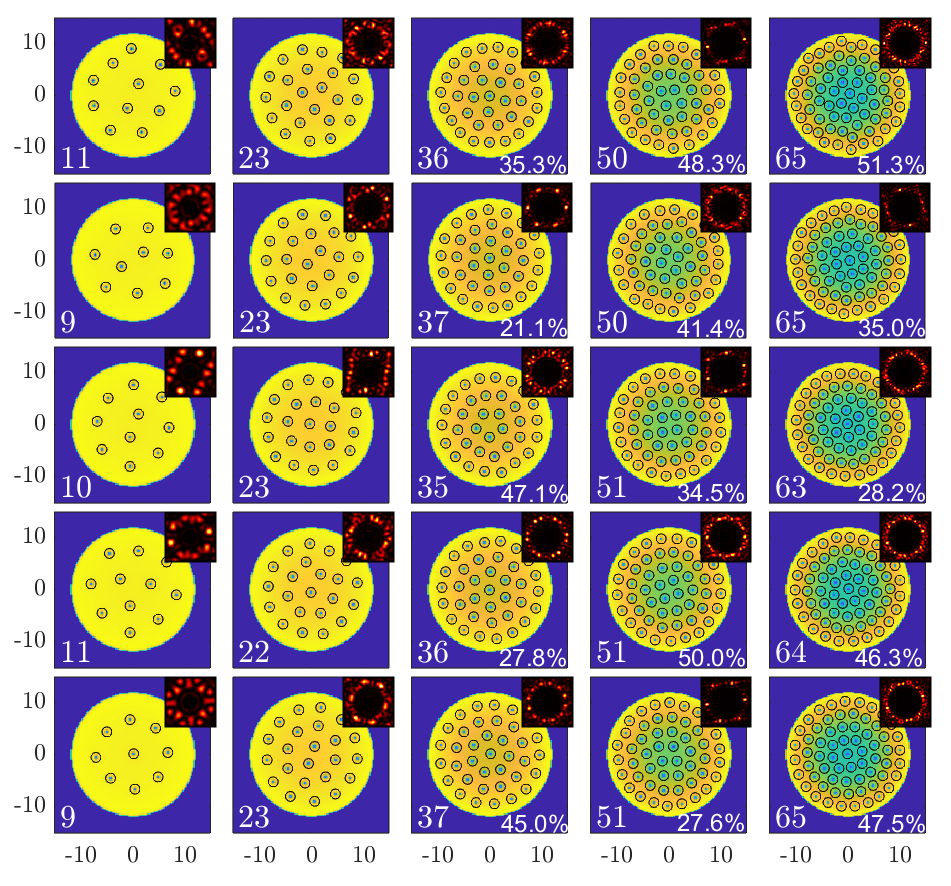}
    
    \caption{\textbf{Post Quench Dynamics I.} We plot the $2D$ condensate density and its vortex-position structure factor $S(\mathbf{k}) = \frac{1}{n_v} \left| \sum_{j=1}^{n_v} e^{-i\mathbf{k}\cdot\mathbf{r}_j} \right|^2$ in the inset for $\mu$-quench (Case I). From left to right, $\Omega = 0.1, 0.2, 0.3, 0.4, 0.5$; from top to bottom, $\tau_{\rm Q} = 50, 100, 200, 300, 400$. The black circles indicate vortices; their total number is given at the bottom left of each panel. On the bottom right of each condensate plot, we mention the corresponding number of defects.}
    \label{fig: grid condensate}
\end{figure}

\begin{figure}[t]
    \centering
    \includegraphics[width=1.0\linewidth]{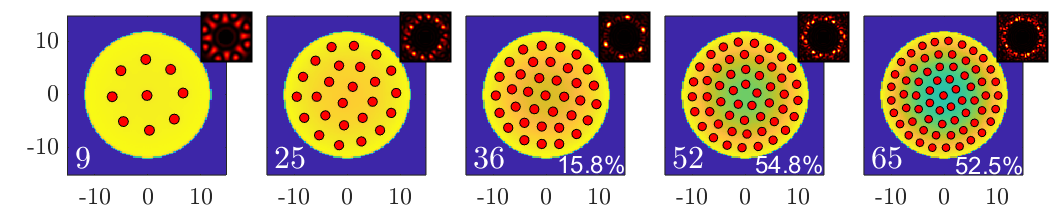}
    \caption{\textbf{Post quench dynamics III.} $2D$ condensate density and its vortex-position structure factor $S(\textbf{k})$ in the inset for $\tau_{\rm Q} = 1000$ and temperature $T = 10^{-7}$ for Case I, with $\Omega = 0.1, \dots, 0.5$ from left to right as in Fig.~\ref{fig: grid condensate}. The vortex number is given at the bottom left and $\rho_{L\text{-def}}$ at the bottom right of panels corresponding to $\Omega \geq 0.3$. The circles represent vortices.}
    \label{fig:grid_condensate_low_temp}
\end{figure}

After the quench, the system settles to the minimal energy state, which depends on the number of defects being formed. The thermal energy of the system and the vortex-vortex repulsion energy drive the vortices towards a certain geometry. Tkachenko~\cite{tkachenko1966vortex} showed that the triangular lattice arrangement of these vortices is the minimal energy state similar to the Abrikosov lattice in Type-II superconductors~\cite{abrikosov1957magnetic, kleiner1964bulk, abrikosov2004nobel, essmann1967direct}. In our system, the geometrical constraint from the trap affects the emergent lattice structure. It is seen from Fig.~\ref{fig: grid condensate} that only for $\Omega = 0.3$, near the bulk center, the arrangement is polygonal, whereas near the periphery it is circular, which is in agreement with Ref.~\cite{Adhikari_2019}. From Fig.~\ref{fig: grid condensate} it can also be seen that, for most of the cases, the internal arrangement is non-polygonal but remains circular on the periphery. This circular arrangement arises from finite-size effects imposed by the trap geometry, with the resulting lattice structure believed to correspond to a natural metastable state of the system. We identify this metastable phase as a vortex glass phase. Such a phase is observed in Type-II superconductors when the vortex lattice transforms from an ordered Bragg glass into a more disordered state once the magnetic field is quenched to $B>B_{\text{ord-dis}}$, with $B_{\text{ord-dis}}$ the critical field of the order-disorder transition~\cite{aragon2019unveiling}. One way to characterize the vortex glass phase is by looking at the vortex lattice-defect density. A lattice-defect (or an $L$-defect) here is defined as the vortex that has a non-sixfold coordination number $(Z)$. The $L$-defect density $\rho_{L\text{-def}}$ is then $\rho_{L\text{-def}} = \left( \frac{N_{\rm Z \neq 6}}{N_{\text{bulk}}}\times 100 \% \right)$, where $N_{\rm Z\neq 6}$ is the number of bulk vortices with $Z\neq6$ and $N_{\rm bulk}$ the total number of bulk vortices (see SM~\ref{Appsec: vortex detection technique}). The metastable state with $\rho_{L\text{-def}}<1\%$~\cite{aragon2019unveiling} is identified as a Bragg glass, whereas for a vortex glass, $\rho_{L\text{-def}}>5\%$~\cite{aragon2019unveiling}. In Fig.~\ref{fig: grid condensate}, we show the calculated $L$-defect density for $\Omega\geq 0.3$ because, below it, the vortex count is simply too low to extract a statistically meaningful coordination number. The diffuse rings and the double peaks in $S(\textbf{k})$ that arise from sub-crystallites are characteristics of the vortex glass phase. With diffused peaks in our $S(\textbf{k})$ graphs, we support this along with the fact that the $L$-defect density in the entire $\Omega(\geq0.3)-\tau_{\rm Q}$ space stays above $15\%$ and we conclude that the system spontaneously nucleates into the vortex glass phase. At this point, to achieve the minimal energy state, we investigated the post quench spatial arrangement for a quench time $\tau_{\rm Q} = 1000$ with a decreased temperature, $T = 10^{-7}$ (shown in Fig.~\ref{fig:grid_condensate_low_temp}).
\begin{figure}[H]

    \includegraphics[width = 1.0\linewidth]{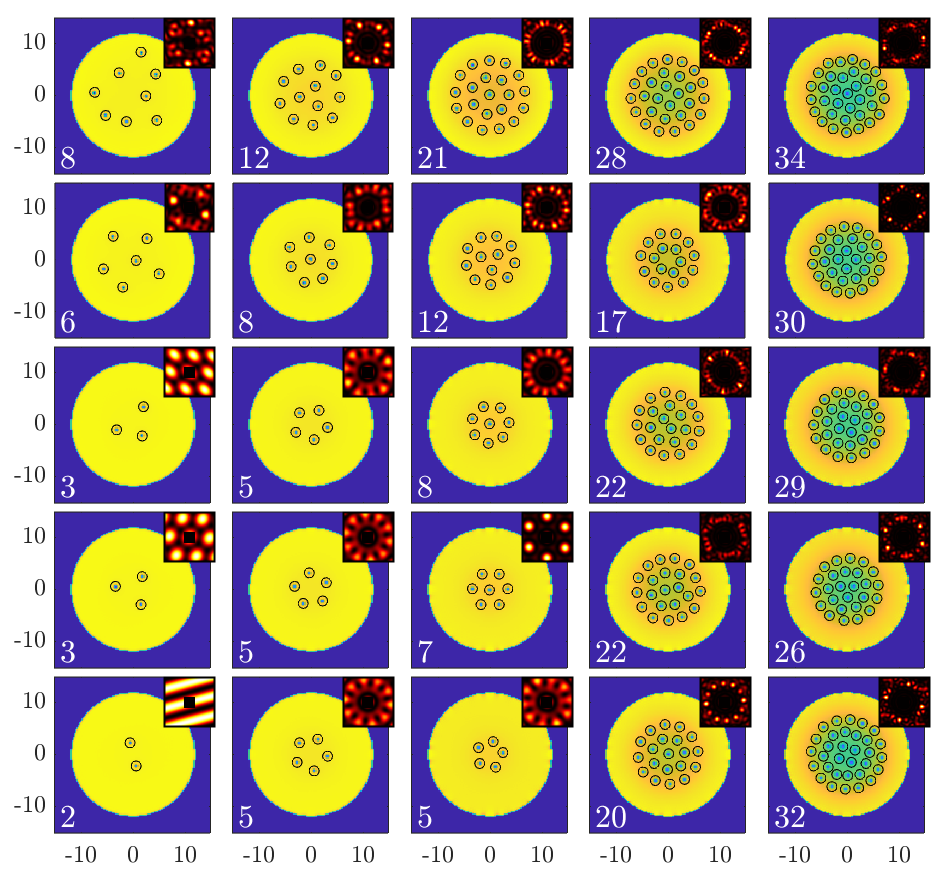}
    
    \caption{\textbf{Post Quench Dynamics II.} We plot the $2D$ condensate density and its vortex-position structure factor $S(\mathbf{k}) = \frac{1}{n_v} \left| \sum_{j=1}^{n_v} e^{-i\mathbf{k}\cdot\mathbf{r}_j} \right|^2$ in the inset for $(\mu,\Omega)$-quench (Case II). From left to right, $\Omega_{\rm f} = 0.1, 0.2, 0.3, 0.4, 0.5$; from top to bottom, $\tau_{\rm Q} = 50, 100, 200, 300, 400$. The black circles indicate vortices; their total number is given at the bottom left of each panel.}
    \label{fig: grid condensate_mu_eff}
\end{figure}

Even in this case the vortex lattice again freezes into a circular geometry at the boundary, indicating the importance of finite size effects. The lattice remains as a topologically frustrated vortex glass. The condensate spontaneously nucleates vortices during the phase transition as it is quenched. The number of vortices remaining sensitive to the chiral bias effect discussed in the main text do not match the number required by the angular momentum. In our hard-wall trap, for a triangular lattice, it is preferred that $n_{\rm v} = 3n(n+1) + 1$, where $n$ is an integer. This condition can be relaxed in the case of a harmonic trap, where the condensate can expand as the angular momentum increases. However, in a hard-wall trap, even if $n_{\rm v}$ is offset by one unit, it acts as an $L$-defect among the lattice sites and alters their geometry. Hence, the vortices cannot settle into a perfect triangular lattice. Similar results were reported in Refs.~\cite{Adhikari_2019, campbell1979vortex, stauffer1968distribution}. A triangular lattice is formed when $n_{\rm v} = 7$(where $n=1$) for $\Omega = 0.3$, $\tau_{\rm Q} = 300$ from Case II as shown in Fig.~\ref{fig: grid condensate_mu_eff}. However, despite having number of vortices, $n_{\rm v} = 37$ where $n = 4$ for $\Omega = 0.3$, $\tau_{\rm Q} = 300$, and $400$ from Case I, no triangular arrangement is seen. This is attributed to the finite size effects from the trap. This indicates that while the macroscopic rotation overrides the initial KZ defect generation, the finite size effects from the trap geometry along with rotation induced chiral bias effects prevent the reliable formation of a perfect lattice. Therefore, achieving a true ground-state crystal requires a mechanism to bypass this topological frustration. An optical pinning potential of the form used in Ref.~\cite{sato2007vortex} can be utilized to drive the lattice into a perfect Abrikosov-lattice arrangement. 

\section{Vortex identification and defect calculation} \label{Appsec: vortex detection technique}
\begin{figure}[H]
    \centering
    \includegraphics[width=1.0\linewidth]{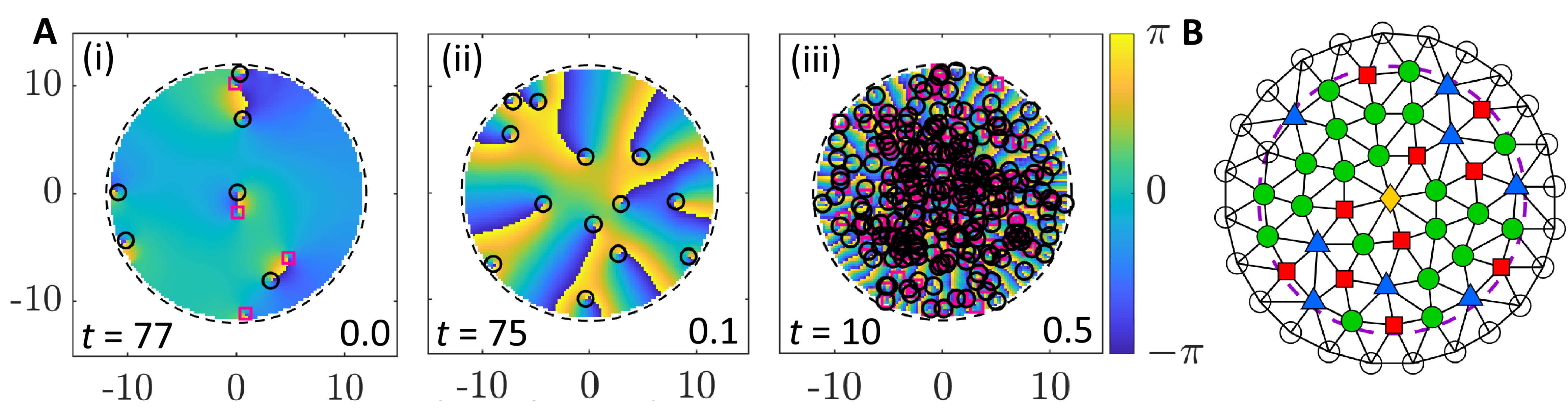}
    \caption{\textbf{(A) Vortex detection and (B) $L$-defect characterization.} \textbf{(A)} We show the phase profile of the condensate for $\Omega = \text{ (i) }0.0, \text{ (ii) }0.1, \text{ and (iii) } 0.5$ for quench time, $\tau_{\rm Q} = 100$ at their corresponding crossover time, $t_{\rm eq}$. The black circles characterize a vortex with a $2\pi$ phase winding and a red square characterizes an anti-vortex with $-2\pi$ phase winding. The black dashed circle corresponds to $0.97R$, beyond which the vortex detection algorithm is not applied. \textbf{(B)} Spatial distribution of quantized vortices connected via Delaunay triangulation to visualize the local coordination number. Nodes are color and shape-coded based on their local coordination number ($Z$): green circles represent $Z=6$, red squares and blue triangles indicate $Z=5 \text{ and }7$ vortices, respectively. Hollow black circles are the boundary vortices and are separated from the bulk by a dashed purple line. Any other coordination number is highlighted with a yellow diamond. The presented lattice is for $\Omega = 0.5$, $\tau_{\rm Q} = 400$. The calculated $\rho_{L\text{-def}}\approx 47.5\%$.}
    \label{Appfig: defect_calculation_delaunay}
\end{figure}

A spontaneously nucleated topological defect in our system is characterized by a quantized circulation of the superfluid velocity, $\oint \textbf{v}(\textbf{r},t) \cdot d\textbf{l} = 2\pi w$, where $w \in \mathbb{Z}$ is the winding number and $w = \pm 1$ is associated with the topological charge of the defects which describes a vortex or an anti-vortex. Given the macroscopic wavefunction $\Psi({\rm x},{\rm y},t) = \sqrt{\rho({\rm x},{\rm y},t)} e^{i\theta({\rm x},{\rm y},t)}$, the topological charge $w$ enclosed by a contour $C$ is fundamentally defined as $w = \frac{1}{2\pi} \oint_C \nabla \theta \cdot d\mathbf{l}$. Oppositely charged vortices carry equal and opposite angular momentum. The associated singularity of the velocity field enforces a vanishing condensate density at the vortex center with the healing length determining the core size. To calculate the number of defects during the quench, topological defects (quantized vortices and antivortices) are identified by computing the phase winding of the complex order parameter around discrete closed loops on the two-dimensional numerical grid. On the discrete spatial grid, the contour integral is mapped to the smallest available closed loop: a $2 \times 2$ elementary plaquette consisting of four adjacent grid points (labeled $1$ through $4$ in counter-clockwise orientation). The discrete topological charge is evaluated as the sum of the nearest-neighbor phase differences around the site: $w = \frac{1}{2\pi} \left( \Delta \phi_{12} + \Delta \phi_{23} + \Delta \phi_{34} + \Delta \phi_{41} \right).$ A site is flagged as containing a topological defect if the total discrete winding takes values $w \approx +1$ or $w \approx -1$. A phase winding is only registered as a physical defect if the local condensate density of the corresponding $2 \times 2$ site is significantly depleted, falling below $50\%$ of the global maximum density of the condensate. We show the phase profiles of the condensate at different rotation frequencies for $\tau_{\rm Q} = 100$ in Fig.~\ref{Appfig: defect_calculation_delaunay}. This algorithm is also useful for detecting any $\Omega$ induced chiral bias in vortex formation even before a well defined condensate is established. As seen in Fig.~\ref{Appfig: defect_calculation_delaunay}A(iii), the algorithm detects a finite number of vortices and antivortices (although spurious and not physical in nature) whose numbers far exceed than what is observed at equilibration time in any case. But plotting a ratio $\frac{|n_+ - n_-|}{n_+ + n_-}$ gives meaningful insights into any chiral bias induced in the system before condensation.
\par
To characterize the emergent crystalline order of the surviving vortices, and strictly distinguish between an Abrikosov lattice and a frustrated vortex glass, the structural geometry of the vortex configurations is analyzed using Delaunay triangulation in Fig.~\ref{Appfig: defect_calculation_delaunay}. From this triangulation, the coordination number $Z$ is computed for each vortex, defined as its total number of nearest neighbors. In an Abrikosov lattice, every vortex in the bulk of the condensate exhibits exactly sixfold coordination ($Z=6$). Therefore, any vortex with $Z \neq 6$ (typically $Z=5$ or $Z=7$) constitutes a structural defect—such as a disclination or an edge dislocation—within the vortex lattice itself. The ratio  $\rho_{L\text{-def}} = \left( \frac{N_{\rm Z \neq 6}}{N_{\text{bulk}}}\times 100 \% \right)$ then gives the density of non-sixfold coordinated vortices. The boundary effects cause geometrical shearing at the edge of the trap. Such vortices are excluded from the $L$-defect analysis, and only those inside the dashed purple line are considered.

\section{Determining the crossover time}
\label{Appsec: Determining the crossover time}
\begin{figure}[H]
\centering
\includegraphics[width=0.5\textwidth]{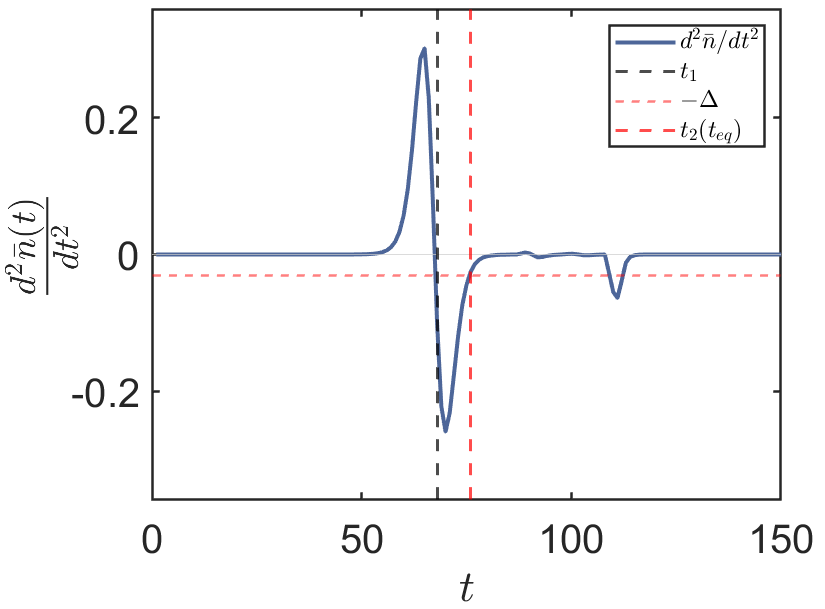}
\caption{\textbf{Determining the crossover time.} The crossover time $t_{\rm eq}$ is systematically determined by evaluating the second derivative of the norm $d^2\bar{n}/dt^2$. We identify the interval $[t_1, t_2]$ where the derivative drops below a threshold $-\Delta$ (with $\Delta$ set to 10\% of the maximum peak). $t_{\rm eq}$ is defined as the upper bound $t_2$. Illustrated here for $\Omega = 0.0$, $\tau_{\rm Q} = 100$.}
\label{fig:picking_the_crossover}
\end{figure}

In this work we have established that the universal KZ scaling of the freeze-out time survives when the rotation is ramped along with the quench, but is lost when the quench is performed in the presence of a pre-existing, fixed rotation. We demonstrated this by calculating $\alpha$ from $\bar{t}_{\rm eq} \propto \tau_{\rm Q}^{\alpha}$, where $\bar{t}_{\rm eq}$ is the crossover time, i.e., the time at which the system crossovers to the regime where it cannot keep adiabatically up with the changing chemical potential. At this instant, the system's dynamics freezes out and is marked as the freeze-out time $\hat{t} \propto \bar{t}_{\rm eq}$. This happens when the instantaneous relaxation time of the system matches the time remaining until the transition. KZ theory predicts that this time $\hat{t}$ measured from either side of the time of crossing the critical point ($t_{\rm c} \mp \hat{t}$) scales universally with $\tau_Q$: $\hat{t} \propto \tau_Q^{z\nu /(1+z \nu)}$, where $t_{\rm c}$ is the time at which the system crosses the critical point $\mu_{\rm c}$, and $\mu_{\rm c} = \mu(t_{\rm c})$. The exponent is $1/2$ the mean field theory. 
However, due to the finite size and interaction effects, the critical point of the condensation can shift~\cite{K_Damle_1996,Giorgini_critical_temp_96,Theory_BEC_Dalfovo_Stefano_review_1999,Liu2020} and is difficult to calculate from first principles~\cite{Liu2020}. 

The standard KZ theory predicts the scaling of the freeze-out time as
\begin{equation}
\hat{t} \;\propto\; \bar{t}_{\rm eq}\mathrel{\mathop:}=(t_{\rm eq}-t_{\rm c})-t_0 \;\propto\; \tau_{\rm Q}^{\,z\nu/(1+z\nu)},
  \label{eq:kzm_scaling_relative_to_tc}
\end{equation}
where $t_{\rm c}$ is the time and $t_0$ is the thermalization time before the quench starts, which we take equal to 10. Then, $t_{\rm c}$ can be given as~\cite{Liu2020}
\begin{equation}
    t_{\rm c} = \tau_{\rm Q} (\mu_{c}-\mu_{i})/(\mu_{\rm f}-\mu_{\rm i}).
    \label{Sm_eq:t_c}
\end{equation}
We establish the KZ scaling in our system by calculating the scaling of the crossover time $\bar{t}_{\rm eq}$ as a function of $\tau_{\rm Q}$ and demonstrate its breakdown in the presence of rotation. To identify the crossover time, we first calculate the time at which the spontaneously formed defects at  freeze-out first equilibrate, i.e., the equilibration time $t_{\rm eq}$~\cite{Chesler15}. We locate the global maximum of $d^2\bar{n}/dt^2$ and identify the subsequent interval $[t_1,t_2]$ over which $d^2\bar{n}/dt^2 \le -\Delta$, with $\Delta$ fixed at $10\%$ of this maximum. We define the equilibration time $t_{\rm eq} \equiv t_2$, as illustrated in Fig.~\ref{fig:picking_the_crossover} for $\Omega = 0.0$ and $\tau_{\rm Q} = 100$. 

To account for the shift in the critical point mentioned above and calculate the crossover time, substituting Eq.~\eqref{Sm_eq:t_c} into Eq.~\eqref{eq:kzm_scaling_relative_to_tc} and writing the KZ scaling as $\bar t_{\rm eq}=C_1\tau_Q^\alpha$ is helpful. The crossover time then can be written as:
\begin{equation}
t_{\rm eq}- t_0 = C_1\,\tau_{\rm Q}^{\alpha} + C_2\,\tau_{\rm Q},
\label{Appeq: teq_two_term_fit}
\end{equation}
where $C_2 = (\mu_{c}-\mu_{i})/(\mu_{\rm f}-\mu_{\rm i})$ is treated as one of the free fitting parameters. This approach is consistent with Refs.~\cite{shinn2025spontaneous,Patra2026-ls}. Table~\ref{tab:crossover_fits} summarizes the fitted parameters for Case I ($\mu$ quench, fixed $\Omega$) and Case II ($(\mu,\Omega)$ quench).

\begin{table}[H]
    \centering
    \caption{Fitted parameters $C_1$, $\alpha$, and $C_2$ for Case I and Case II at different final rotation frequencies. The values in parentheses denote the uncertainty in the last reported digits.}
    \label{tab:crossover_fits}
    \renewcommand{\arraystretch}{1.25}
    \setlength{\tabcolsep}{7pt}
    \small
    \begin{tabular}{c ccc @{\qquad} ccc}
        \toprule
        & \multicolumn{3}{c}{\textbf{Case I} ($\mu$ quench)}
        & \multicolumn{3}{c}{\textbf{Case II} ($( \mu,\Omega )$ quench)} \\
        \cmidrule(lr){2-4}
        \cmidrule(lr){5-7}
        $\Omega/\Omega_{\rm f}$
        & $C_1$ & $\alpha$ & $C_2$
        & $C_1$ & $\alpha$ & $C_2$ \\
        \midrule
        $0.0$ & $7.113(90)$   & $0.485(4)$  & $0.004(3)$
              & $7.113(90)$  & $0.485(4)$  & $0.004(3)$ \\
              
        $0.1$ & $7.311(53)$   & $0.476(2)$  & $-0.004(2)$
              & $7.168(57)$  & $0.482(2)$  & $0.008(2)$ \\

        $0.2$ & $7.512(148)$  & $0.468(6)$  & $-0.055(6)$
              & $6.897(40)$  & $0.490(2)$  & $0.004(2)$ \\

        $0.3$ & $10.297(386)$ & $0.344(10)$ & $-0.036(4)$
              & $6.454(57)$  & $0.503(3)$  & $-0.002(3)$ \\

        $0.4$ & $12.581(421)$ & $0.218(8)$  & $-0.015(2)$
              & $6.147(114)$ & $0.510(6)$  & $-0.002(6)$ \\

        $0.5$ & $13.234(462)$ & $0.112(8)$  & $-0.005(1)$
              & $5.775(47)$  & $0.519(3)$  & $-0.004(2)$ \\
        \bottomrule
    \end{tabular}
\end{table}

Moreover, the system does not contain a well-defined number of topological defects until a well-defined condensate forms that necessarily lies outside the freeze-out domain $[-\hat{t}, +\hat{t}]$. Thus, $t_{\rm eq}$ is also the natural time scale to measure the density of topological defects~\cite{Chesler15}.

\bibliographystyle{apsrev4}
\bibliography{references}